\documentclass{aa}  

\usepackage{graphicx}
\usepackage{txfonts}
\usepackage{xspace}
\usepackage{hyperref}

\newcommand{\swift}{{\it Swift}\xspace}
\newcommand{\xmm}{{\it XMM-Newton}\xspace}

\newcommand{\ROSAT}{{\it ROSAT}\xspace}
\newcommand{\ero}{\hbox{eROSITA}\xspace}
\newcommand{\srg}{{\it SRG}\xspace}
\newcommand{\cxo}{{\it Chandra}\xspace}

\newcommand{\src}{{XMMU\,J010429.4$-$723136}\xspace}
\newcommand{\sxp}{{SXP\,164}\xspace}

\newcommand{\ergcm}[1]{$\times 10^{#1}$ erg cm$^{-2}$ s$^{-1}$}
\newcommand{\oergcm}[1]{$10^{#1}$ erg cm$^{-2}$ s$^{-1}$}

\newcommand{\oergs}[1]{$10^{#1}$ erg s$^{-1}$}
\newcommand{\hcm}[1]{$\times 10^{#1}$ cm$^{-2}$}
\newcommand{\ohcm}[1]{$10^{#1}$ cm$^{-2}$}

\newcommand{\Halpha}{{H${\alpha}$}\xspace}
\newcommand{\Hbeta}{{H${\beta}$}\xspace}
\newcommand{\ltsima}{$\buildrel < \over \sim$}
\newcommand{\lsim}{\lower.5ex\hbox{\ltsima}}
\newcommand{\gtsima}{$\buildrel > \over \sim$}
\newcommand{\gsim}{\lower.5ex\hbox{\gtsima}}

\usepackage{etoolbox}

\begin{document}

   \title{SRG/\ero discovery of 164\,s pulsations from the SMC Be/X-ray binary \src}
   
   \titlerunning{Discovery of \sxp's X-ray pulsations}

   \author{S. Carpano\inst{1}
          \and
          F. Haberl\inst{1}
          \and
          C. Maitra\inst{1}
          \and
          M. Freyberg\inst{1}
          \and
          K. Dennerl\inst{1}
          \and
          A. Schwope\inst{2}
          \and
          A. H. Buckley\inst{3}
          \and 
          I. M. Monageng \inst{3,4}
          }

   \institute{Max-Planck-Institut f\"{u}r extraterrestrische Physik, Gie{\ss}enbachstra{\ss}e 1, 85748 Garching, Germany\\ \email{scarpano@mpe.mpg.de}
   \and Leibniz Institut f\"ur Astrophysik Potsdam, An der Sternwarte 16, D-14482 Potsdam, Germany
   \and South African Astronomical Observatory, P.O. Box 9, Observatory, Cape Town 7935, South Africa
   \and Department of Astronomy, University of Cape Town, Private Bag X3, 7701 Rondebosch, South Africa
   }

   \date{Received April XX, 2021; accepted XX XX, 2021}

 
  \abstract
   {The Small Magellanic Cloud (SMC) hosts many known high-mass X-ray binaries (HMXBs), and all but one (SMC X-1) have a Be companion star. Through the calibration and verification phase of \ero on board the Spektrum-Roentgen-Gamma (\srg) spacecraft, the Be/X-ray binary \src was in the field of view during observations of the supernova remnant, 1E0102.2$-$7219, used as a calibration standard.}
   {We report timing and spectral  analyses of \src based on three \ero observations of the field, two of which were performed on 2019 November 7-9, with the third on 2020 June 18-19. We also reanalyse the OGLE-IV light curve for that source in order to determine the orbital period.}
   {We performed a Lomb-Scargle periodogram analysis to search for pulsations (from the X-ray data) and for the orbital period (from the OGLE data). X-ray spectral parameters and fluxes were retrieved from the best-fit model.}
  {We detect, for the first time, the pulsations of \src at a period of $\sim$164\,s, and therefore designate the source as \sxp. From the spectral fitting, we derive a source flux of $\sim$1$\times 10^{-12}$ erg~s$^{-1}$~cm$^{-2}$ for all three observations, corresponding to a luminosity of $\sim$4$\times 10^{35}$~erg~s$^{-1}$ at the distance of the SMC. Furthermore, reanalysing the OGLE light curve, including the latest observations, we find a significant periodic signal that we believe is likely be the orbital period; 
at 22.3\,d, this is shorter than the previously reported values. The \swift/XRT light curve, extracted from two long monitorings of the field and folded at the same period, suggests that a modulation is also present in the X-ray data.}
   {}


   \maketitle
%

\section{Introduction}
\label{sec:intro}
High-mass X-ray binaries (HMXBs) are binary systems composed of an early-type star and a compact object that is either a neutron star or a black hole (occasionally a white dwarf). Be or B[e]-type stars are a subset of B-type stars where one or more Balmer emission lines are found in the optical spectrum. They are believed to possess an equatorial decretion disc, possibly causing regular X-ray outbursts at periastron passage of the compact object due to enhanced mass accretion \citep[see e.g.][for a review]{Reig2011}. A large number of such systems were found in the Small Magellanic Cloud (SMC). The latest comprehensive catalogue was published by \cite{Haberl2016} and collects 121 high-confidence HMXBs (the vast majority with Be companion stars). For about half of the sample, X-ray pulsations were discovered with periods ranging from a fraction of a second to a few thousand seconds. Pulse periods of the 62 Be/X-ray binaries have a bimodal distribution peaking at around 10\,s and 250\,s \citep{Haberl2016}. For the other sources, pulse periods are still not reported, one of them being \src, identified as source number 132 in Table~A.1 of this latter publication. The source was already mentioned as source number 3285 in the previous \xmm catalogue of X-ray point sources in the SMC \cite[][their Table~5]{Sturm2013}.

\src was confirmed as a Be star (more precisely of type B1\,V) by \citet{McBride2017} based on spectra recorded with the Anglo-Australian Telescope on 2012 July 7–8 in the wavelength range [4025--4775]\,\AA. Large X-ray variability had previously been reported by \citet{Maggi2013} comparing a series of \swift/XRT observations, where the luminosity ranged from 2.8 to 12.4$\times 10^{35}$~erg~s$^{-1}$ on MJD 56627.09 to 56643.31 (2013 December), with a previous \xmm observation on MJD 55149.1 (2009 November), where only an upper limit a factor of $\sim$400 lower could be extracted. The source is not reported in the RXTE catalogue of SMC pulsars from \cite{Galache2008} based on 9\,yr of SMC survey (starting in 1997), but was detected by \cxo in 2002 August with a luminosity of 1.4$\times 10^{35}$~erg~s$^{-1}$ \citep{Rajoelimanana2011}.

The Optical Gravitational Lensing Experiment  (OGLE) light curve of \src in the $I$-band is shown in \citet{McBride2017} as source XMM~3285, covering the epoch from MJD 55346 to MJD $\sim$57400 (2010 May to 2016 January). The source was in a low state for $\sim$600\,d and then increased towards a stable higher state ($\Delta I$=1.25\,mag). A period of 29.75\,d was found in the data covering the short time interval MJD 55650--56100 (and after some detrending). This is somewhat lower than the strong modulation of 37.15$\pm$0.02\,d claimed by \citet{Rajoelimanana2011} and associated later by \citet{Schmidtke2013}, with aliasing of a possible shorter period modulation of 0.972\,d. This last short period was interpreted as non-radial pulsations from the Be star. We note that in both works from \cite{Rajoelimanana2011} and \cite{Schmidtke2013}, the source is erroneously named SXP\,707 from a period discovered in the {\cxo} data at 707\,s. That period is, in reality, the satellite dithering period, with the source located at the rim of a CCD, moving in and out of the detector \citep[as already clarified by ][]{Haberl2016}.

In this paper, we report the discovery of  X-ray pulsations for \src with a period of 164\,s and rename the source \sxp following the  terminology first proposed by \citet{Coe2005} (where SXP stands for Small Magellanic Cloud X-ray Pulsar, followed by the pulse period in seconds to three significant figures). The discovery was briefly announced by \citet{Haberl2019}.  We also report the detection of a strong signal at 22.3\,d in the OGLE-IV light curve that we interpret as the possible orbital period.

\section{Observations and data reduction}

X-ray observations were performed during the calibration and verification phase of \ero \citep{Predehl2021}, from UTC 2019 November 07 17:13:18 to 2019 November 08 09:53:18 (ObsID 700001) and 2019 November 09 02:41:38 to 2019 November 09 19:21:38 (ObsID 700003), each observation for a total exposure of $\sim$60\,ks. A third observation occurred half a year later from 2020 June 18 19:43:25 to 2020 June 19 06:00:41 ($\sim$37\,ks, ObsID 710000). During the first observation, the source was far off-axis (at an angle of 29.8$'$) and was covered by only four cameras, telescope modules TM1,\,4,\,5,\,6 (for TM3, as only a small fraction of the PSF lies in the field of view, the data are not included in the analysis). Instead, the source was located in the field of view of all seven cameras in the second and third observations (with an off-axis angle of 26.8\arcmin\  and 28.4\arcmin, respectively).
The data were reduced using the \ero Standard Analysis Software System pipeline \citep[\texttt{eSASS};][]{2021arXiv210614517B} version \texttt{eSASSusers\_201009}. The software determines good time intervals, corrupted events and frames, and dead times, and masks bad pixels, projects the photons onto the sky, and applies pattern recognition and energy calibration. Source and background extraction regions were defined as an ellipse and an elliptical annulus, respectively. The regions were centred on the \xmm coordinates  of the source (RA, DEC)= (01:04:29.42, -72:31:36.5), derived in \cite{Sturm2013}. For the observation 700001, an elliptical nearby background region was selected instead as the source is observed far off-axis.

Optical spectroscopy of \src was undertaken on 2019-11-24 using the Robert Stobie Spectrograph \citep[RSS, ][]{Burgh2003} on the Southern African Large Telescope \citep[SALT, ][]{Buckley2006} under the transient follow-up program. The PG2300 VPH grating was used, which covered the spectral region 6100--6900\,\AA\ at a resolution of 1.2\,\AA. A single 1200 s exposure was obtained, starting at 19:40:08 UTC. 

The X-Ray variables OGLE Monitoring (XROM) system provides real-time photometry ($I$-band) of optical counterparts of X-ray sources\footnote{\url{http://ogle.astrouw.edu.pl/ogle4/xrom/xrom.html}} located in the fields observed in the OGLE-IV survey \citep{Udalski2008}, with roughly daily sampling. To date, data available for \sxp cover the period from MJD 55346 (2010-05-30) to MJD 58870 (2020-01-21).

\section{Data analysis and Results}

\subsection{\ero}
\subsubsection{X-ray light curves and pulse period search}
\label{sec:xray_time}

\begin{figure}
   \centering
   \resizebox{\hsize}{!}{\includegraphics{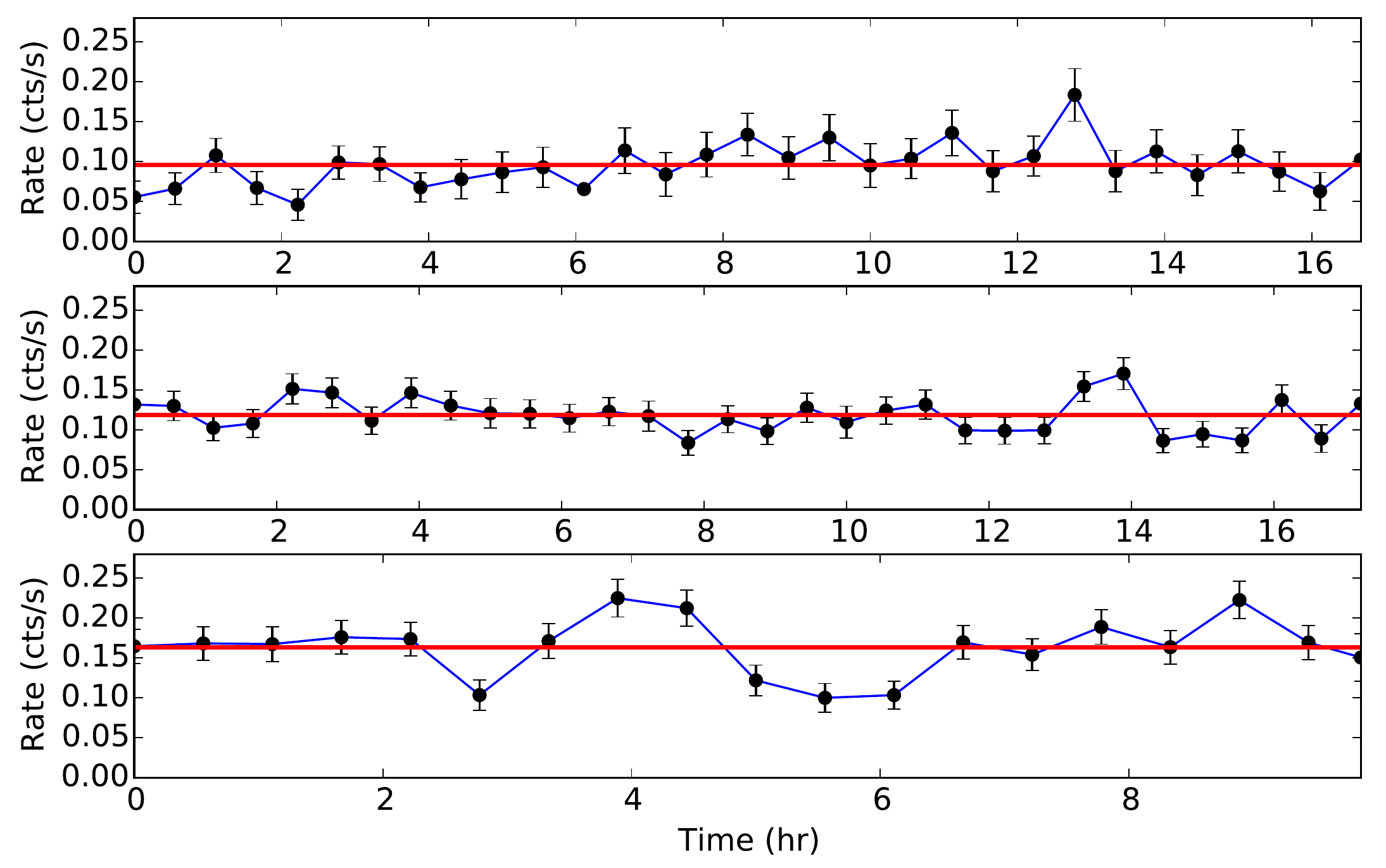}}
   \caption{\ero light curve of \sxp, during ObsID 700001 (top), 700003 (middle) and 710000 (bottom), rebinned at 2000\,s, extracted in the 0.2--5.0\,keV band. The red lines indicate the mean values of 0.10, 0.12, and 0.16 cts/s for the three respective observations.}
              \label{fig:sxp164_LC}             
\end{figure}

The background-subtracted and vignetting-corrected light curves of \sxp from the three observations performed during the calibration phase are shown in Fig.~\ref{fig:sxp164_LC}. These were produced using the \texttt{eSASS} task \texttt{srctool} in the energy band 0.2--5.0\,keV and with a bin size of 2000\,s.
The mean count rates, shown by the red lines, are 0.10, 0.12, and 0.16\,cts\,s$^{-1}$, for the three respective observations. The portion of the light curves where all available cameras are operating simultaneously has almost the same length for the first two observations ($\sim$17\,hr) and is much shorter for the third one ($\sim$10\,hr). No large variability is observed during the individual observations.

\begin{figure}
   \centering
   \resizebox{\hsize}{!}{\includegraphics{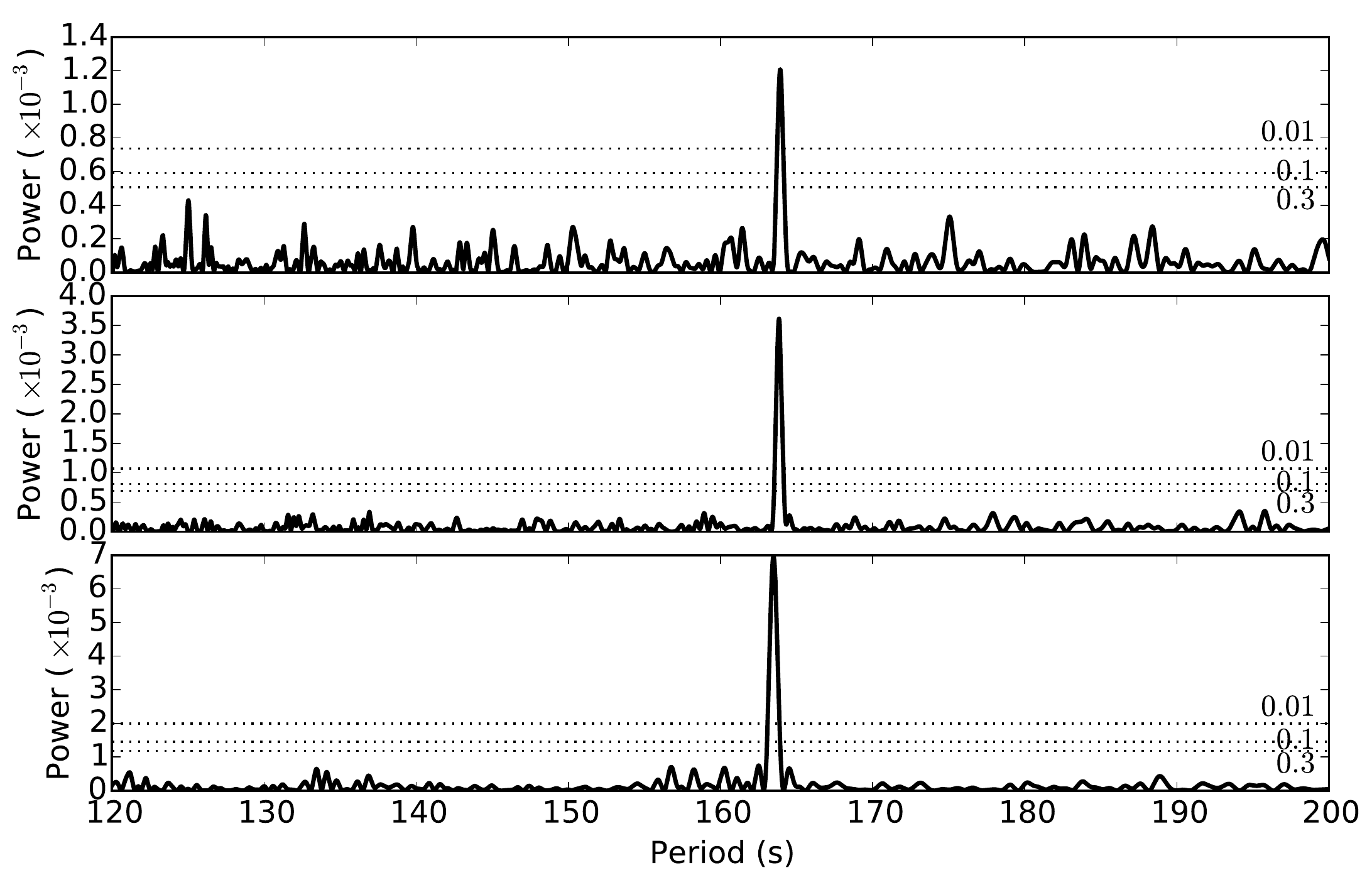}}
   \caption{Lomb-Scargle periodogram for the three \ero observations (top: ObsID 700001, middle: 700003 and bottom: 710000),
   using light curves binned at 2\,s. The best periods are 163.93\,s,  163.85\,s, and 163.48\,s, respectively.}
              \label{fig:sxp164_periodogram}             
\end{figure}

The pulse period search is performed using a Lomb-Scargle analysis \citep{Lomb1976, Scargle1982} in the energy band 0.2--5\,keV (as the signal is stronger than in the 0.2--10\,keV band) and in the period range 120--200\,s (a larger period interval was initially used during the first screening).  Figure~\ref{fig:sxp164_periodogram} shows the periodogram corresponding to the three observations, performed on background-subtracted light curves binned at 2\,s. Strong peaks are found with maxima at 163.93\,s, 163.85\,s, and 163.48\,s, respectively. 
The confidence levels (given at 68$\%$, 90$\%,$ and 99$\%$) are derived using the block-bootstrap method as explained in \cite{Carpano2017}. The analysis consists first in simulating 1000 light curves, each by splitting the original light curves into blocks of $\sim$650\,s and randomly shuffling the blocks, and then retrieving the corresponding periodogram maxima. 
We also performed a joint period search on the first two data sets as the values are very close, and we find the best period at 163.89\,s.
To estimate the uncertainty on the pulse periods, we carried out Monte Carlo simulations, similarly to what is done in \cite{Gotthelf1999}, and generated a set of 1000 light curves for each observation.  The block-bootstrapped light curves are used to recreate the noise while the signal is represented by a sine function whose amplitude is retrieved from the folded light curves (and the period, assumed to be constant, from the maxima of the periodograms). In that way, the significance of the peak in the periodograms derived from the simulated data is about the same as with the original light curves. The corresponding measured periods are well represented by a normal distribution and the mean values are very close to the input periods (163.926\,s, 163.848\,s, and 163.489\,s). The standard deviation of the distribution is taken as the 1\,$\sigma$ uncertainty in the period and is 0.042\,s, 0.030\,s, and 0.051\,s, for the three observations, respectively. The uncertainty on the period retrieved by merging the first two data sets is 0.007\,s (for a mean value of 163.897\,s).

\begin{figure}
   \centering
   \resizebox{\hsize}{!}{\includegraphics{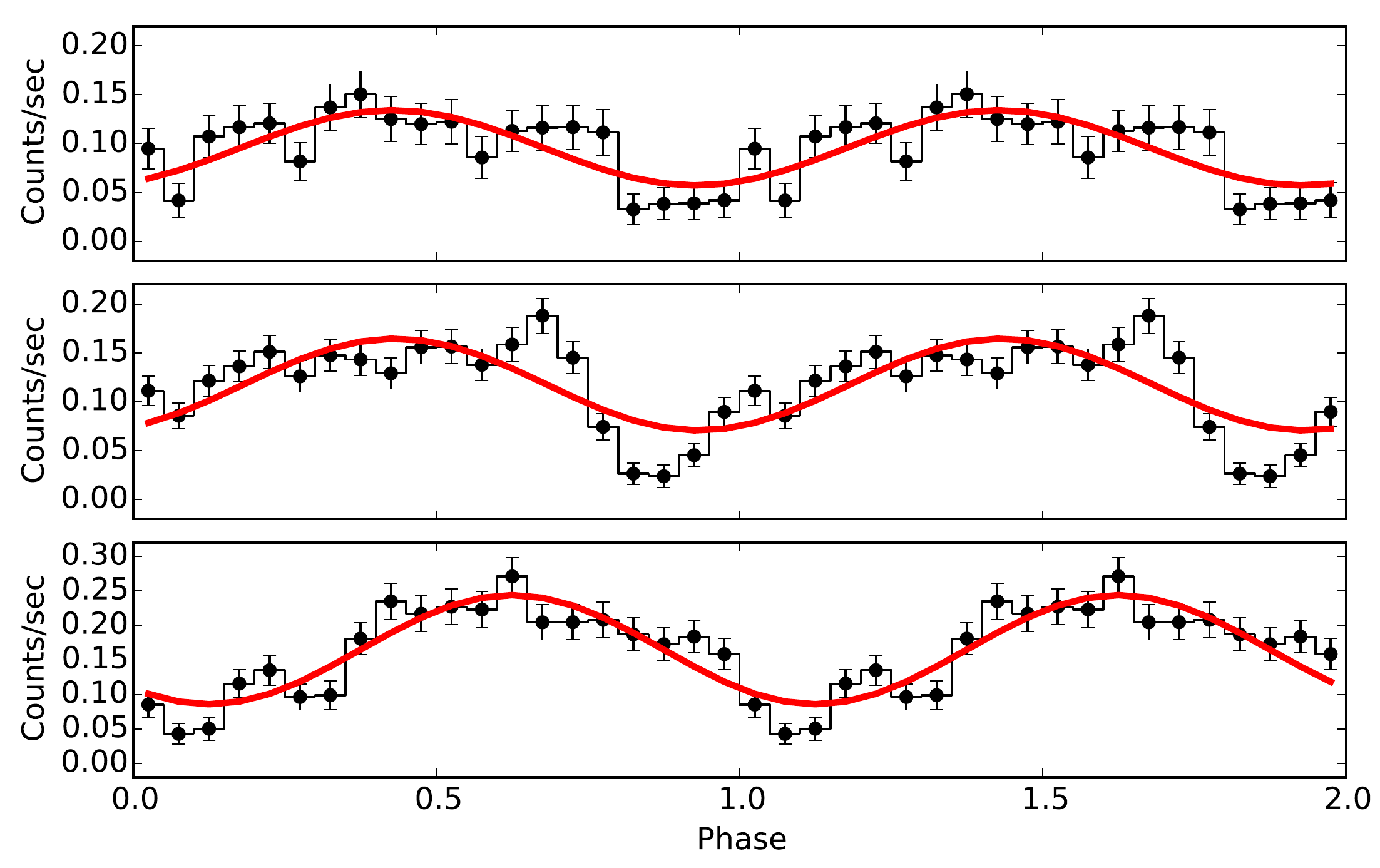}}
   \caption{Folded \ero light curves of \sxp during ObsID 700001 (top), 700003 (middle), and 710000 (bottom), extracted in the 0.2--5.0\,keV band. The best-fit sine function is overlaid. Phase 0 corresponds to the start of the first observation. The light curves from the first two observations are folded with the common period (163.89\,s).}
              \label{fig:folded_LC_sxp164}             
\end{figure}

Figure~\ref{fig:folded_LC_sxp164} shows the folded light curves of \sxp with the best fit sine functions overlaid. For the first two observations, they are folded at their common period (163.89\,s), while the third one is folded at its own best period. The light curves are phase connected with phase `0' being the start of the first observation. The pulse profile of the second observation looks less sinusoidal than that of the third one (the first one is more noisy as less photons are available). The reduced chi-square of the sine fit (defined as $\chi^2_{red}=\sum\limits_{i} \frac{(obs_i-fit)^2}{\sigma_i^2} /\textrm{d.o.f.}$  with $obs_i$ being the folded rate, $fit$ the sine function, $\sigma_i$ the rate errors, and d.o.f. the degrees of freedom) is 1.7, 5.4, and 2.3 for the three respective observations. On the other hand, the pulsed fraction (defined as $PF=\frac{max-min}{max+min}\times100$ (\%), with $max$ and $min$ being the maximum and minimum of the folded light curves) is 64\%, 78\%, and 73\% for the first, second and third observation, respectively.

\begin{figure}
   \centering
   \resizebox{\hsize}{!}{\includegraphics{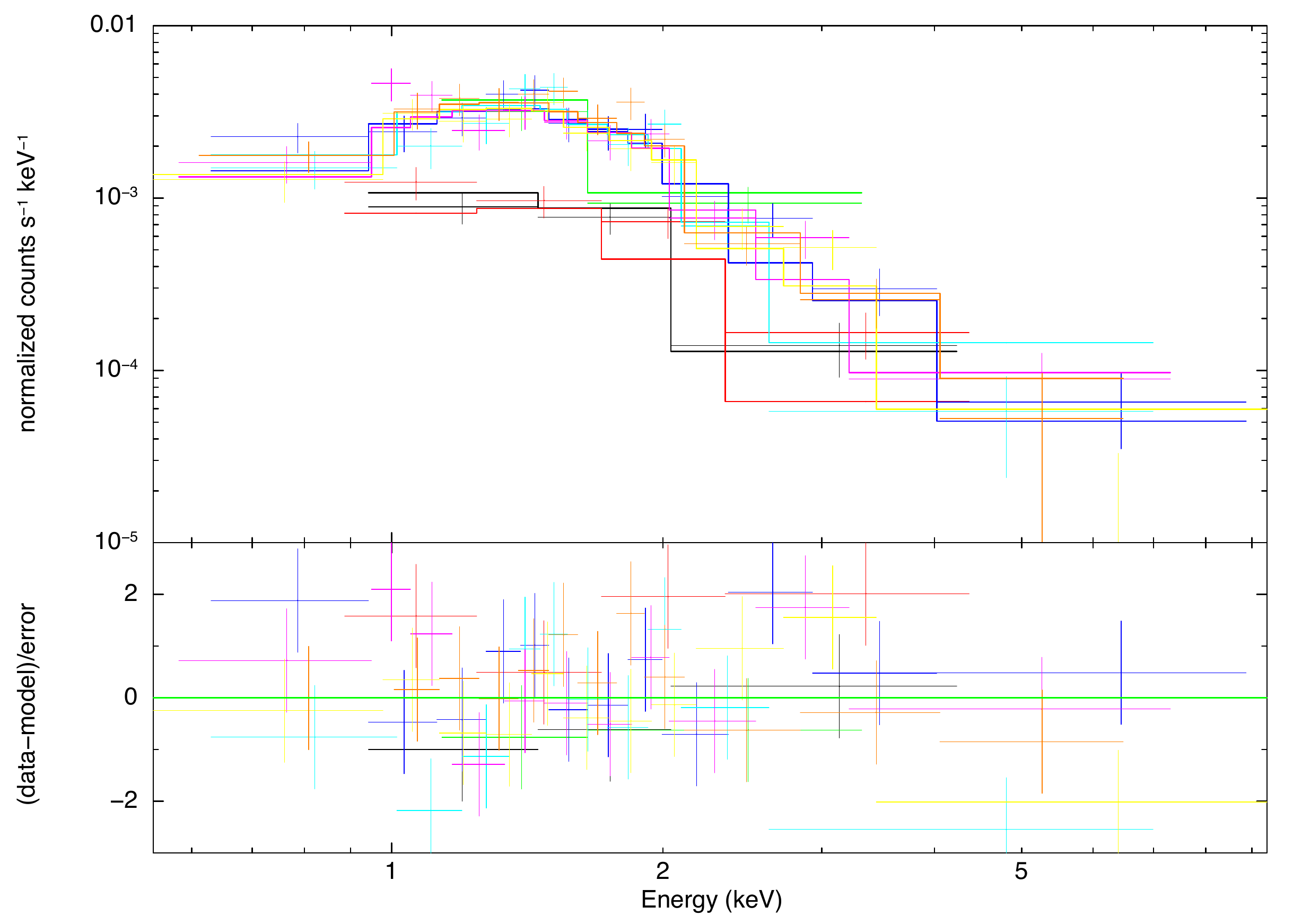}}
   \resizebox{\hsize}{!}{\includegraphics{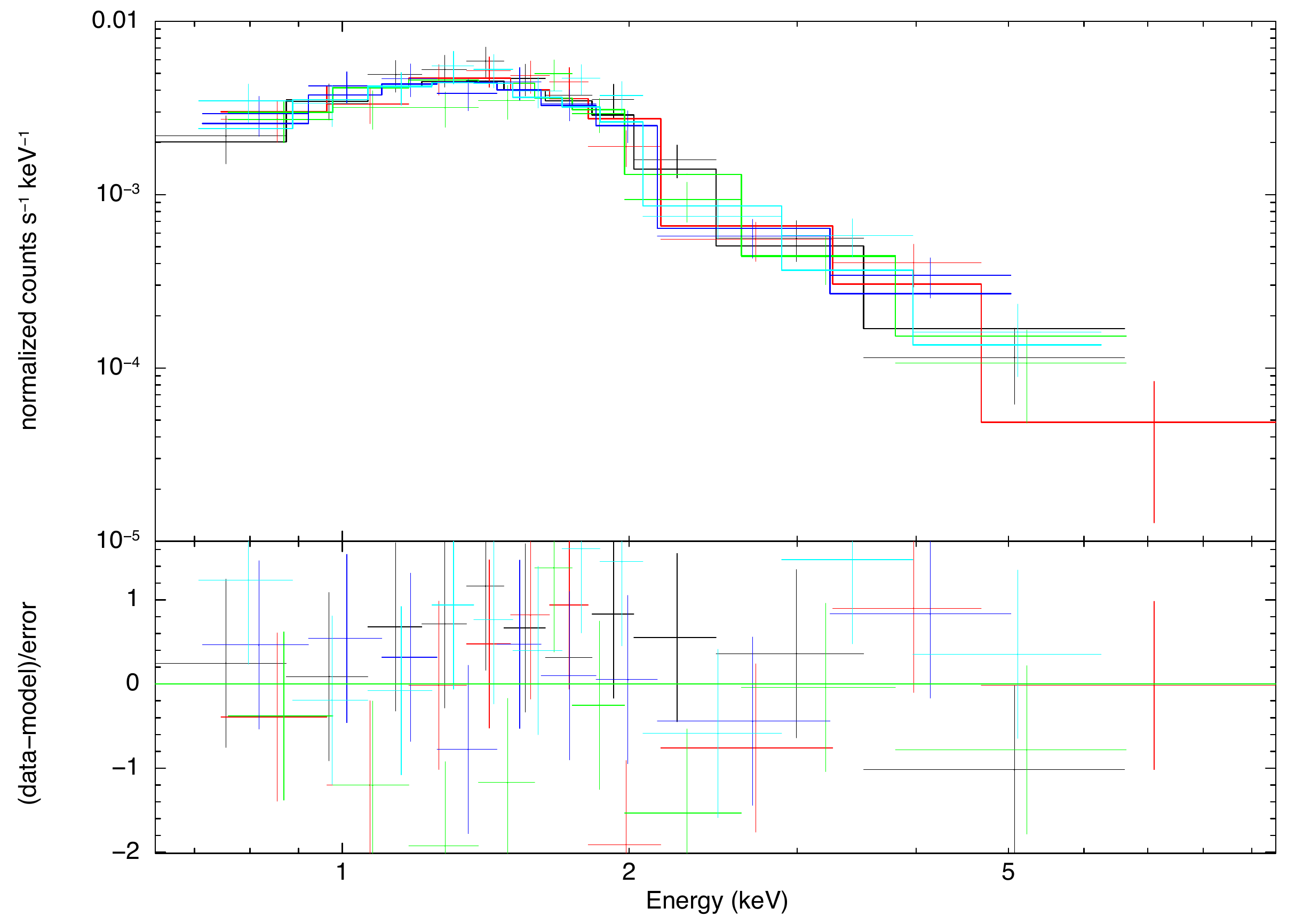}}
   \caption{Simultaneous spectral fit of \sxp using an absorbed power-law model (\texttt{TBabs*TBvarabs*power}), and the corresponding residuals for ObsID 700001 (3 cameras) and 700003 (5 cameras) at the top and ObsID 710000 (5 cameras) at the bottom.}
   \label{fig:spec_sxp164}             
\end{figure}

\subsubsection{Spectral analysis}
\label{sec:xray_spec}

For all three observations, the \ero spectra of \sxp were fitted with \texttt{PyXspec}\footnote{\url{https://heasarc.gsfc.nasa.gov/xanadu/xspec/python/html/index.html}} in the full 0.2--10\,keV energy band, and for the first two observations simultaneously (fitting the data for the two observations separately would lead to similar spectral parameters but with larger uncertainties). The spectra for all cameras covering the sources \citep[excluding the light-leak cameras TM5 and TM7,][]{Predehl2021} were fitted using an absorbed power law and two absorption components (\texttt{TBabs*TBvarabs*power}). \texttt{TBabs} and \texttt{TBvarabs} are the  Tuebingen-Boulder interstellar medium (ISM) absorption models as defined in \cite{Wilms2000}. The first absorption component has a fixed Galactic N$_\textrm{H}$ value of 5.36$\times$10$^{20}$\,cm$^{-2}$ \citep{Dickey1990} and the second N$_\textrm{H}$ is left free, while the abundance is fixed to 1.0 for He and 0.2 for Z>2 \citep{Russell1992}. The second X-ray absorption component arises in the ISM of the SMC or close to the source.

For the first two observations, only the normalisation constant for the spectra of different observations was left free. The results of the spectral fit are shown in Fig.~\ref{fig:spec_sxp164} together with their residuals,  for ObsID 700001 and 700003 (top), and 710000 (bottom). The values of the spectral parameters as well as the observed and unabsorbed fluxes and luminosities \citep[assuming a distance of 60.6\,kpc,][]{Hilditch2005} are shown in Table~\ref{tab:specral_params}. The observed fluxes are in the range of those reported by \cite{Maggi2013}, 0.66 to 2.87\ergcm{-12}, from the \swift observations mentioned in Sect.~\ref{sec:intro}, where count rates were converted into 0.3--10\,keV fluxes assuming an absorbed power-law model with N$_{\mathrm{H}}$=1\hcm{21} and $\Gamma$=1. 
The power-law index derived here is consistent with what is retrieved for other SMC Be/X-ray binaries. An average value of 0.93 was reported in \cite{Haberl2008} (90$\%$ of the values being between 0.71 and 1.27) based on \xmm observations of 20 Be/X-ray binaries with luminosities above \oergs{35}, for which spectral fitting was possible.

Neither the flux nor the spectral shape vary significantly from observation to observation, indicating the source is relatively constant on both short timescales of a few days and over a six-month period (see also the fluxes derived from the \ero survey in Sect.\,\ref{sec:long_LC}).

\begin{table}
  \caption[]{Spectral parameters resulting from the fit to the \ero spectra during ObsID 700001 and 700003, performed simultaneously, and 710000, and the corresponding fluxes (absorbed and unabsorbed) and luminosities (in the 0.2--10\,keV band and assuming a distance of 60.6\,kpc)}.

\begin{tabular}{lccc}
    \hline
    Observation & 700001 & 700003 & 710000 \\
    \hline
    $\ensuremath{N_{\mathrm{H}}}$ (\texttt{TBabs})\tablefootmark{a}  &  0.0536 (frozen) & - & -\\
    $\ensuremath{N_{\mathrm{H}}}$ (\texttt{TBvarabs})\tablefootmark{a}  &  0.63$^{+0.26}_{-0.22}$ & - & 0.45$^{+0.29}_{-0.26}$\\
    $\Gamma$ & 0.93$^{+0.25}_{-0.22}$ & - & 0.74$^{+0.28}_{-0.25}$\\
    $\chi^2$  & 79.8 & -& 38.0\\
    d.o.f. & 70 & - & 47\\
    Flux$^a$ & 1.19$^{+0.27}_{-0.17}$& 0.87$^{+0.13}_{-0.12}$ & 1.50$^{+0.21}_{-0.25}$\\
    Unabs. Flux \tablefootmark{a}  & 1.27 & 1.00 & 1.56\\
    Unabs. Lum. \tablefootmark{b}  & 5.56 & 4.08 & 6.85\\
    \hline
    \end{tabular}
    \tablefoot{Units: 
       \tablefoottext{a}{\ohcm{22}}; 
       \tablefoottext{b}{\oergcm{-12}}; 
       \tablefoottext{c}{\oergs{35}}
    }
    \label{tab:specral_params}
\end{table}

\subsection{SALT-RSS spectrum}

\begin{figure}
   \centering
   \resizebox{\hsize}{!}{\includegraphics{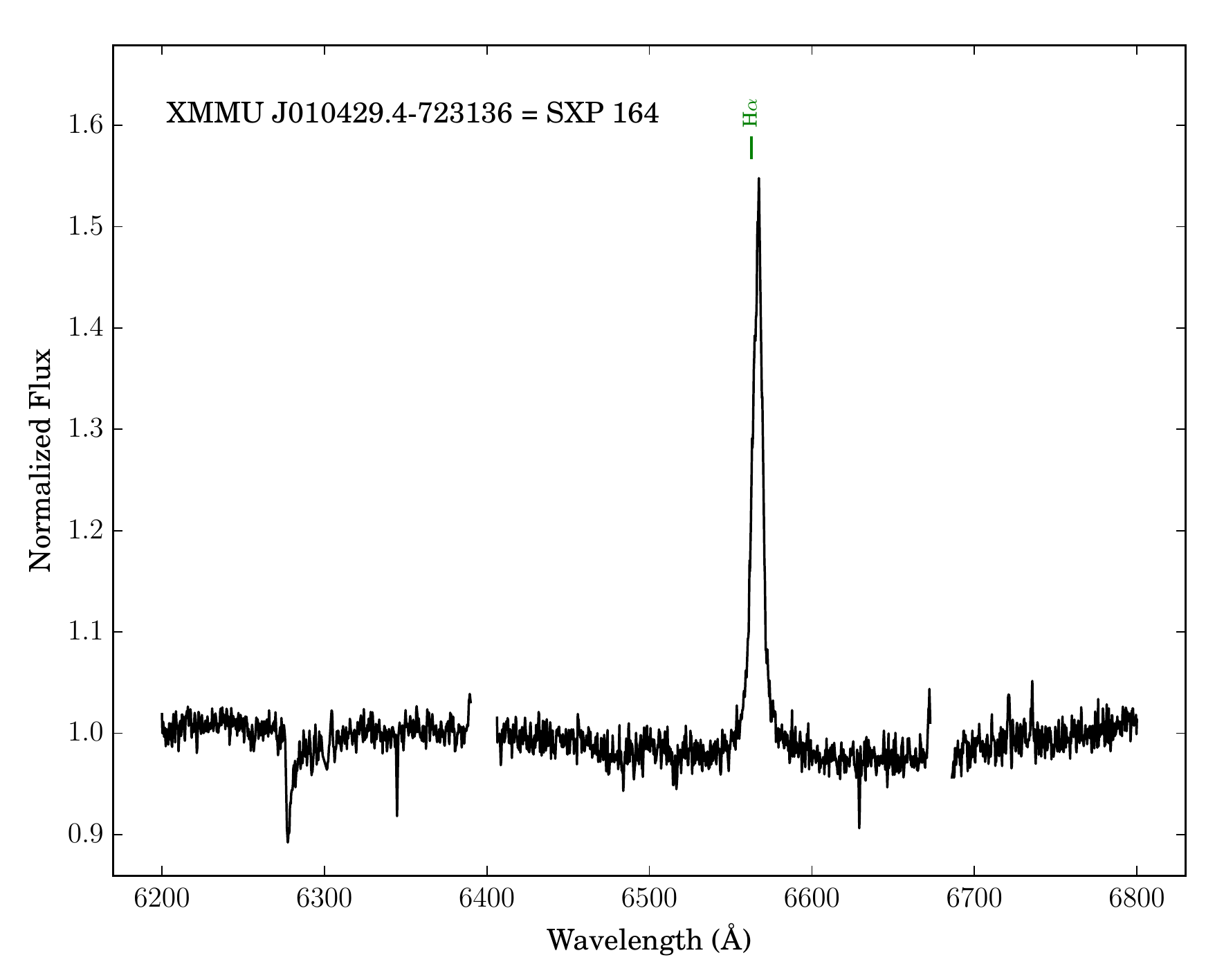}}
   \caption{Optical spectrum of \src obtained on 2019-11-24 using the Robert Stobie Spectrograph on SALT. The rest wavelength of the \Halpha line is marked.}
   \label{fig:SALT}             
\end{figure}

Figure~\ref{fig:SALT} shows the optical spectrum of \src taken with SALT-RSS. 
The spectrum is dominated by a single-peaked, slightly asymmetric \Halpha emission line, with a steeper gradient on the redward side. The measured line properties are: the equivalent width EW = -5.12$\pm$0.14 \AA, full width at half maximum FWHM = 7.224$\pm$0.023 \AA,\ and central position of the \Halpha line = 6566.358$\pm$0.028 \AA\ (corrected to the Solar System barycentre). 

We note that the identification of the source as a Be star by \cite{McBride2017}  was based on the blue part of the optical spectrum (4025 to 4775\,\AA). There, the Balmer lines, including the \Hbeta line, all appeared in absorption, as is observed for many other systems. Here, the \Halpha line, observed in emission, confirms the nature of the source as a Be star. Furthermore, precise measurements of the line EW and FWHM are reported for the first time.

\subsection{OGLE-IV data and orbital period search}

\begin{figure}
   \centering
   \resizebox{\hsize}{!}{\includegraphics{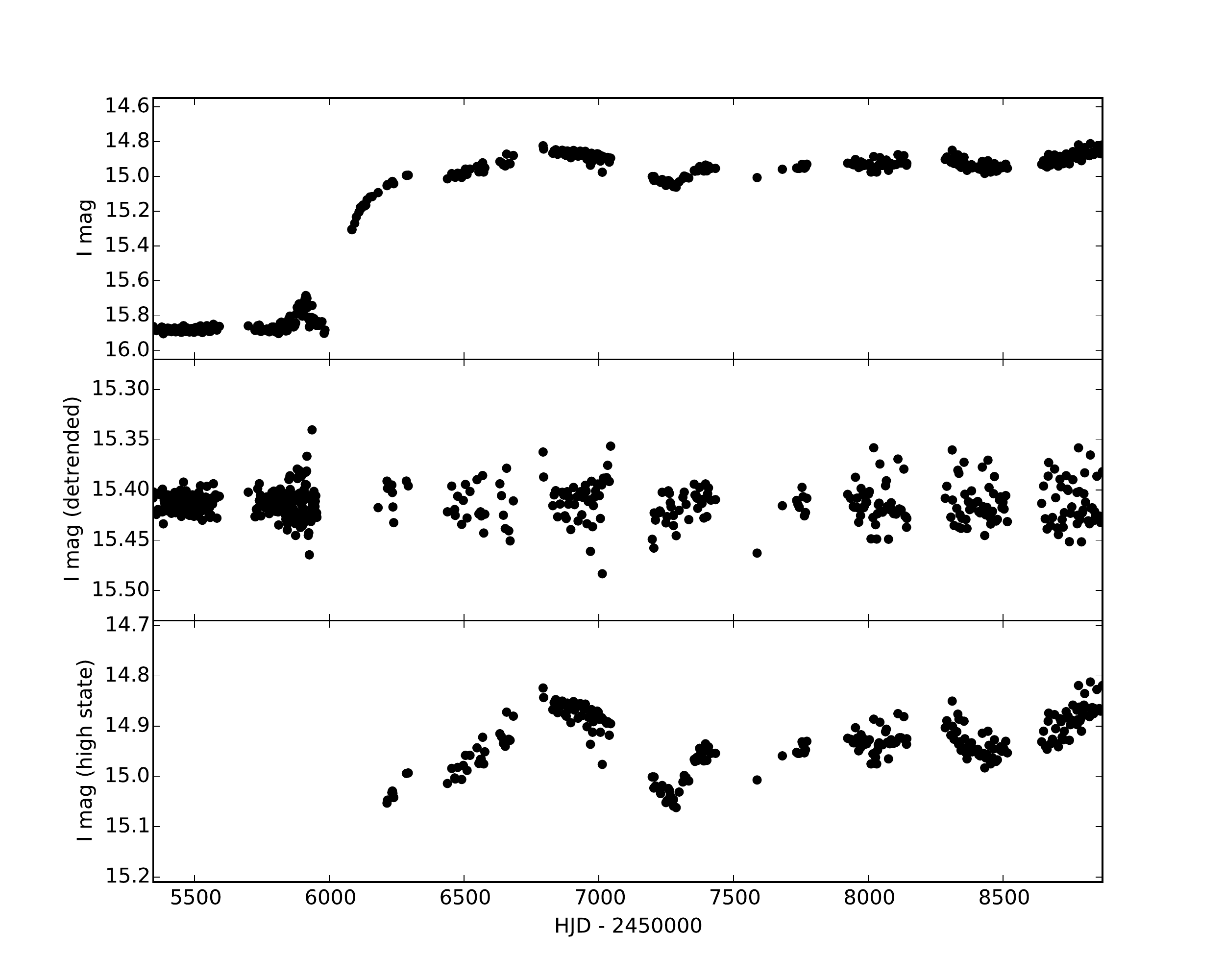}}
   \caption{OGLE-IV light curve of \sxp in the \textit{I} band: full light curve (top), detrended (middle), high state only (bottom).}
    \label{fig:OGLE_LC}             
\end{figure}

The upper part of Fig.\,\ref{fig:OGLE_LC} shows the OGLE-IV light curve in the \textit{I} band, as provided in the XROM data archive. The observations cover the period ranging from 2010 May to 2020 January. The source is first at low flux level for the first $\sim$600\,d and then increases into a high state over a timescale of  $\sim$1\,yr (from 2011 November to 2012 October), where it stays until the end of the observations. In an attempt to look for a periodic signal possibly associated with the orbital period, we detrended the light curve using a hamming window function (with \texttt{Python} tool \texttt{PyAstronomy.pyasl.smooth}) and removed the data points where the flux abruptly increases. The resulting light curve is shown in the middle of Fig.\,\ref{fig:OGLE_LC}. Finally, the bottom part shows the portion of the light curve with the source in the bright state only, representing  a large fraction of the data (with respect to its fainter state).

\begin{figure}
   \centering
   \resizebox{\hsize}{!}{\includegraphics{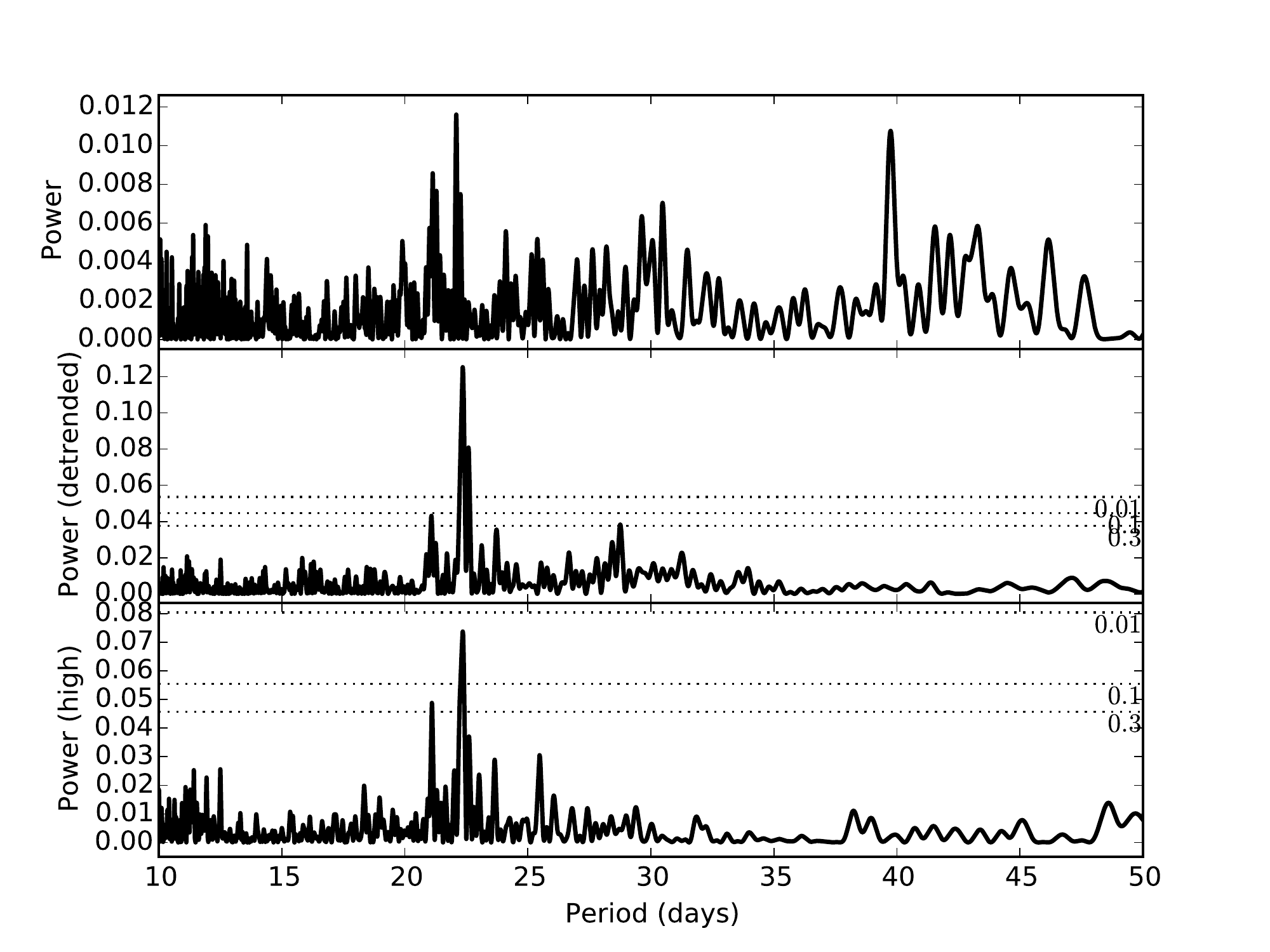}}
   \caption{Lomb-Scargle periodogram of the OGLE full (top), detrended (middle), and high-state-only (bottom) light curves showing a clear peak at 22.35\,d for the detrended and high-state-only light curves (22.09\,d in the full light curve). The confidence levels (given at 68\%, 90\% and 99\%) are derived using the block-bootstrap method).}
    \label{fig:OGLE_LS}             
\end{figure}

On each light curve, a period search was performed using the same Lomb-Scargle analysis as in Sect.\,\ref{sec:xray_time}, in the period range from 10 to 50\,d, and the results are shown in Fig.\,\ref{fig:OGLE_LS}.  For the full light curve,  the maximum peak is found at 22.09\,$\pm$0.04\,d but is not very pronounced (confidence levels using the bootstrap method cannot be used because of the flux increase).
For the detrended or high-flux-only light curves, the best period is found at 22.35$\pm$0.11\,d in both cases and the signal is significant at least at 90\% confidence level. We note that for the detrended light curve, a change of the window width of the smooth algorithm would slightly change the shape of the periodogram, but the best period value would still be consistent within the error bars and the significance of the peak would still be above 99\%. Furthermore, no other significant peak is found between 50\,d and 150\,d for the detrended and high-flux-only light curves.

\begin{figure}
   \centering
   \resizebox{\hsize}{!}{\includegraphics{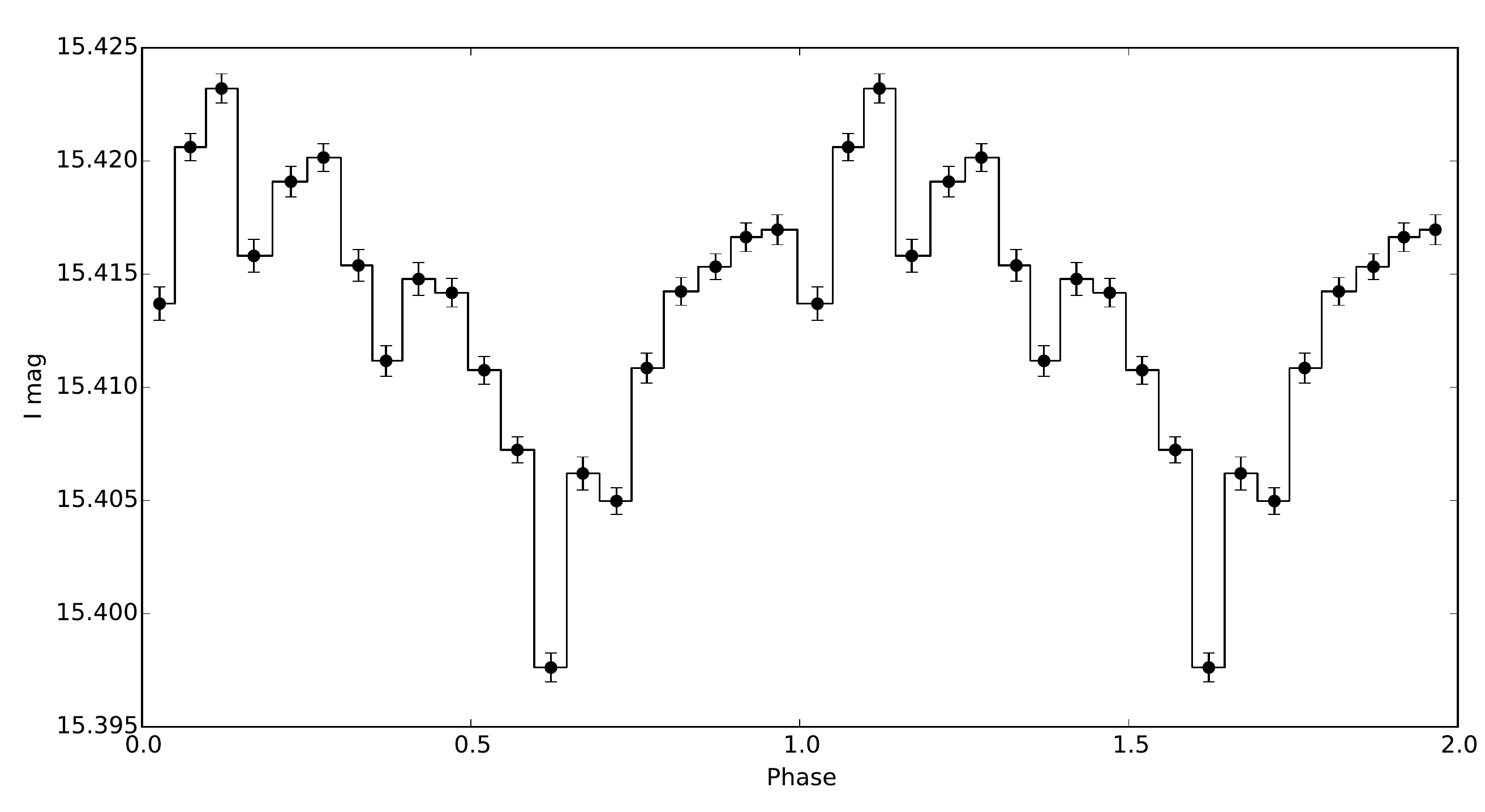}}
   \caption{Detrended OGLE light curve folded at the best period, the shape being symmetric and having more of a sinusoidal than a FRED-like profile.}
    \label{fig:OGLE_folded}             
\end{figure}

The detrended light curve folded at the best period is shown in Fig.~\ref{fig:OGLE_folded}, where Phase `0' corresponds to the start of the observation (MJD: 55378.44102). The shape is symmetric and has more of a sinusoidal than a FRED(Fast Rise Exponential Decay)-like profile \citep[see][for a comparison of both profiles]{Bird2012}. FRED-like profiles for orbital modulations are expected in Be-X-ray binaries, which would be caused by the neutron star disturbing the circumstellar disc of the Be star.

\subsection{Long-term X-ray light curve }
\label{sec:long_LC}
\begin{figure}
   \centering
   \resizebox{\hsize}{!}{\includegraphics{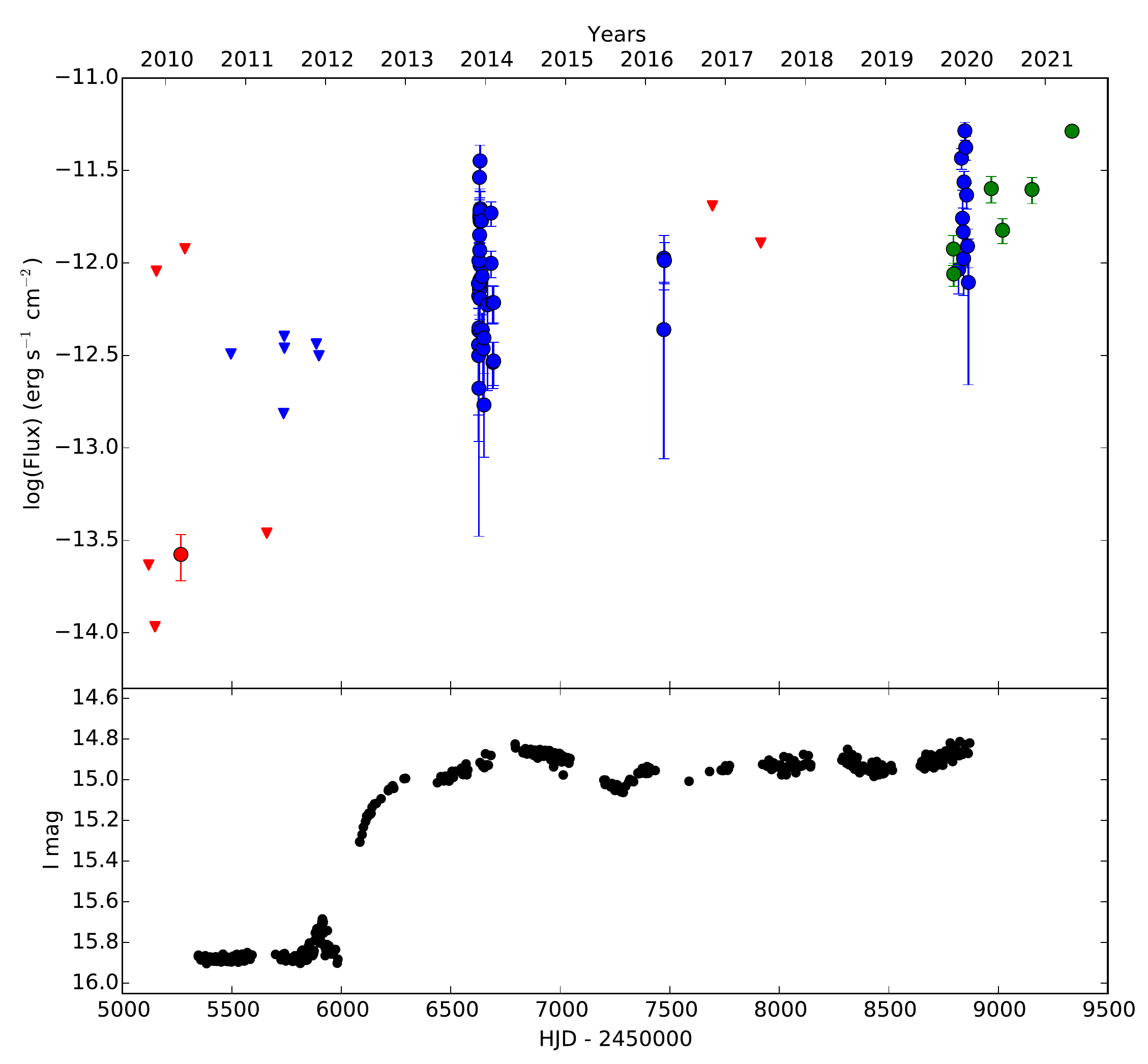}}
   \caption{Long-term X-ray light curve of SXP164 (top) during the OGLE monitoring (bottom). Dots represent measured flux, while upper limits are shown as triangles. 
   \xmm (pointed+slew), \swift and \ero fluxes are shown in red, blue, and green, respectively.
   Flux values before 2019 are derived using the ULS.}
    \label{fig:long_LC}             
\end{figure}

Figure~\ref{fig:long_LC} shows the long-term X-ray light curve of SXP164 in the time interval covered by the OGLE-IV monitoring (shown again at the bottom). The measured 0.2--12\,keV fluxes (dots) and the upper limits (triangles) are derived from both pointed and slew \xmm and \swift observations using the Upper Limit Server (ULS\footnote{\url{http://xmmuls.esac.esa.int/upperlimitserver}}). The values are derived assuming a power-law spectrum with N$_{\mathrm{H}}=1\times10^{21}$\,cm$^{-2}$ and a spectral index $\Gamma$=1, as in \cite{Maggi2013}. \cxo observed the source only once, in 2002 August, where the flux was 3.2 $\times 10^{-13}$ erg\,cm$^2$\,s$^{-1}$ \citep{Rajoelimanana2011}. The 0.2--10\,keV fluxes from the most recent \swift observations (2019 December 1 to 2020 January 15), which are not included in the ULS, are shown as well and assume the same spectral model (see below). In addition to the pointed observations described in this paper, the source was observed during the \ero all-sky surveys eRASS1 (in 2020 April-May),  eRASS2 (in 2020 November) and eRASS3 (in 2021 May) with count rates of 0.269$\pm$0.044\,cts\,s$^{-1}$,  0.226$\pm$0.043\,cts\,s$^{-1}$ , and 0.546$\pm$0.065\,cts\,s$^{-1}$ (average over successive $\sim$40\,s scanning observations separated by gaps of  $\sim$4\,hr). The count rates are converted into fluxes using the same scaling factor as for the pointed observations. The resulting \ero fluxes, as derived in Sect.\ref{sec:xray_spec} and from the surveys, are shown in  Fig.~\ref{fig:long_LC} in green.

\begin{figure}
   \centering
   \resizebox{\hsize}{!}{\includegraphics{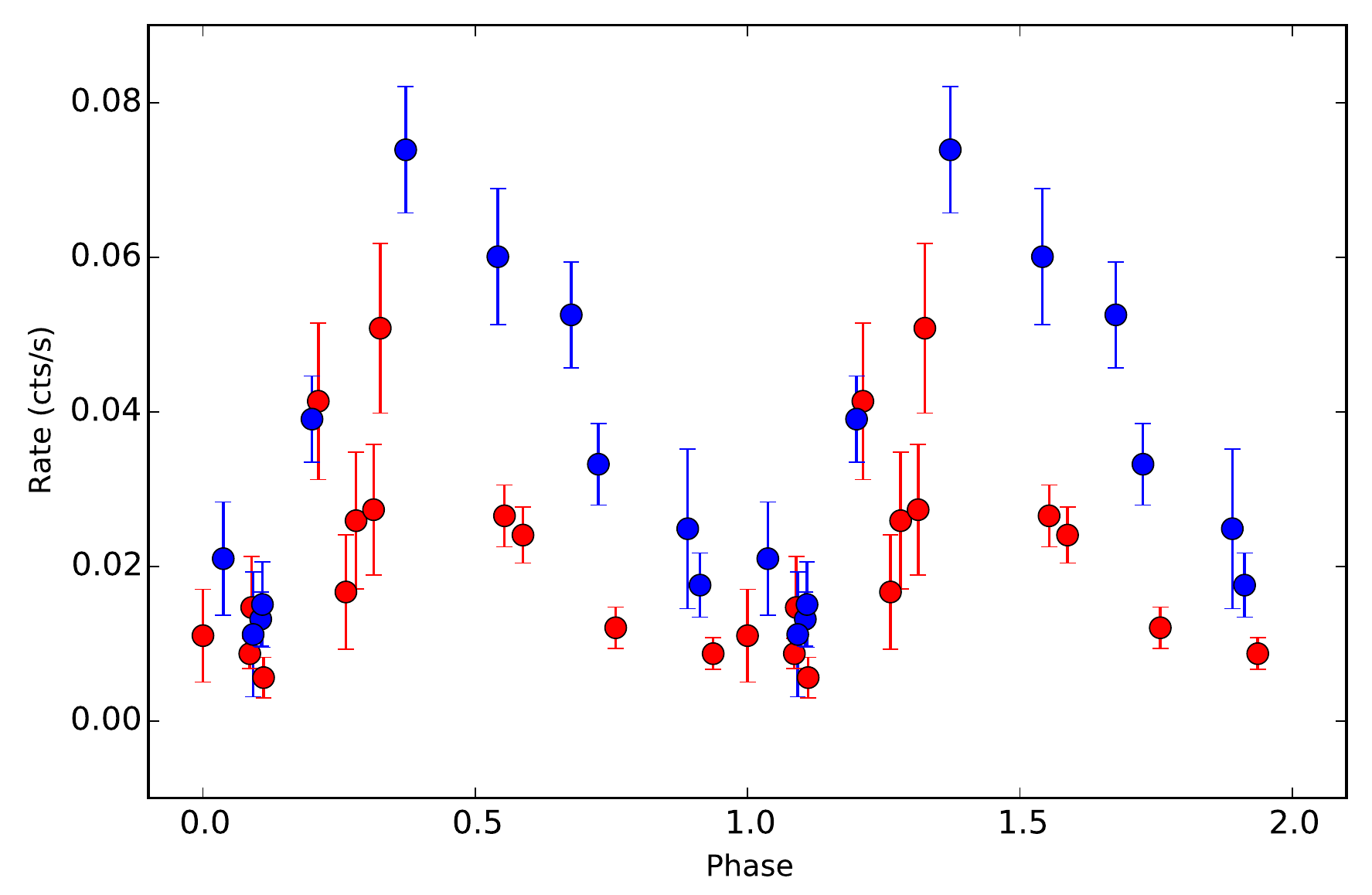}}
   \caption{Vignetting-corrected light curve of \sxp from two sets of \swift/XRT observations (2013 November to 2014 February in red and 2019 December to 2020 January in blue) folded at the orbital period (22.35\,d).}
    \label{fig:folded_orbital}             
\end{figure}

The long-term light curve indicates the possibility  that the source became bright simultaneously in the X-ray and the \textit{I} bands, i.e. from late 2012. Some variability (a factor of $\sim$10) is nevertheless present in the high X-ray state. 
To check if the X-ray variability observed after 2013 could be related to the binary orbit (Type-I outbursts) we analysed two sets of \swift/XRT observations: the first performed from 2013 November 30 (ObsID: 00033042001) to 2014 February 07 (00033042038) with SXP\,65.8 as target (only with off-axis angles $<12'$), and the second one running from 2019 December 01 (00012207001) to 2020 January 15 (00012207011), with \src as the target.
The 0.3--10\,keV vignetting-corrected count rates for these observations were retrieved using the online tool of the UK \swift Science Data Centre\footnote{\url{https://www.swift.ac.uk/user_objects}}, which is described in \cite{Evans2007} and \cite{Evans2009}. 
 The resulting light curve folded at the orbital period (22.35\,d) is shown in Fig.~\ref{fig:folded_orbital}. These two data sets clearly show large-amplitude modulations that are phase-connected and could be very well explained by the orbital motion. The suggested orbital period is therefore also visible in the X-rays, although a longer monitoring with \swift/XRT would be necessary to increase the significance of the modulation.

\section{Discussion}
We report the discovery of the pulse period (164\,s) and possibly the orbital period (22\,d) of a known Be/X-ray binary in the SMC. 
We had a close look at the existing archival X-ray data prior to \ero to check whether or not the pulse period could have been detected. In the archival \ROSAT and \xmm data, from which the detections are reported by \cite{Haberl2000}\footnote{\href{https://vizier.u-strasbg.fr/viz-bin/VizieR-5?-ref=VIZ60b77da2393bf&-out.add=.&-source=J/A\%2bAS/142/41/table2&recno=185}{https://vizier.u-strasbg.fr/viz-bin/VizieR-5?-ref=VIZ60b77da2393bf\&-out.add=.\&-source=J/A\%2bAS/142/41/table2\&recno=185}} and in \cite{Rosen2016} \footnote{\href{https://www.ledas.ac.uk/rfs/LEDAS/3XMM-DR7/catfiles/0601213201/C0601213201EPX000SRCSUM0014.HTM}{https://www.ledas.ac.uk/rfs/LEDAS/3XMM-DR7/catfiles/0601213201/C0601213201EPX000SRCSUM0014.HTM}}, the source is too faint, while for the \swift observations, the exposures are too short. Only in the unique \cxo observation from 2002 were sufficient photons available, but no period other than that of the spacecraft dithering can be detected.

As for many other Be/X-ray binaries belonging to that galaxy, the source is highly variable in the X-rays with a ratio F$\textrm{max}$/F$\textrm{min}\gtrsim$750 with F$\textrm{max}\sim3.8\times10^{-12}$ and F$\textrm{min}\lesssim $5\ergcm{-15} (upper limit from a \xmm observation performed in October 2000). Those values are in line with what is reported by \cite{Haberl2016} in their Fig.\,5 for the other SMC Be/X-ray binaries.
Following the analysis of the \swift/XRT observations described in Sect.\,\ref{sec:long_LC}, most of the variability could be explained by an orbital modulation at a period of 22\,d, which is consistent with the period derived here from the analysis of the OGLE-IV data.
We note that \cite{Schmidtke2013} studied the MACHO and OGLE-II data available at the time and find that only a few seasons presented evidence for any periodicity. In OGLE-II data, only the first season shows a prominent period at P=0.972\,d or its alias at 37.1\,d.
A peak near the 39.7\,d period is also visible in our periodogram, but only for the full undetrended light curve.

According to the well-known Corbet diagram \citep{Corbet1986}, an empirical power-law relationship between the pulse period P$_\textrm{s}$ (in seconds) and the orbital period P$_\textrm{o}$ (in days) was derived for Be/X-ray binaries based on the observed values, with $e$ being the orbital eccentricity:
\begin{equation}
  P_\textrm{o}=10(1-e)^{-3/2}P_\textrm{s}^{0.5}
  \label{equ:corb}
.\end{equation}

A more recent and updated version of this empirical relationship was proposed by \cite{Yang2017} as:
\begin{equation}
  P_\textrm{o}=12.066 P_\textrm{s}^{0.425}
  \label{equ:yang}
.\end{equation}

\begin{figure}
   \centering
   \resizebox{\hsize}{!}{\includegraphics{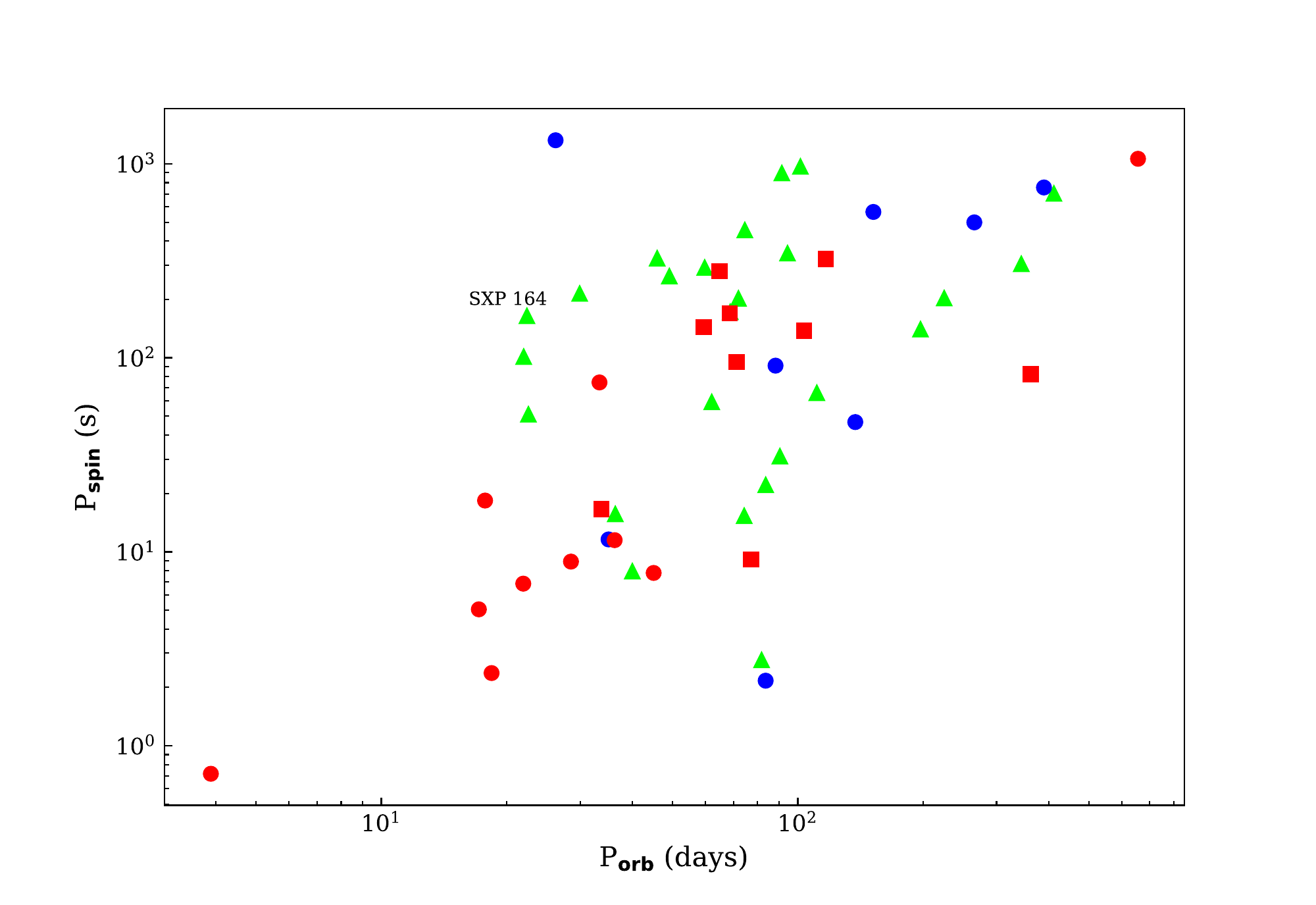}}
   \caption{Spin period vs. orbital period for HMXBs in the SMC, as presented in \cite{Haberl2016}, including \sxp. Orbital periods found with different methods are marked with different symbols: full orbit solution (red circles), X-ray outbursts (red squares), optical light curve (green triangles). Blue circles mark systems with orbital periods consistently derived from X-rays and optical.}
    \label{fig:corbet}             
\end{figure}

This relationship can be explained for systems being in a state of quasi-equilibrium in which the Alfv\'{e}n radius (the radius where the inflowing matter couples to the magnetic field lines of the neutron star) and corotation radius (radius where the spin angular velocity of the neutron star is equal to the Keplerian angular velocity) are equal on average. However, on the Corbet diagrams, a large scatter is observed in both papers, as is also shown by \cite{Haberl2016}, indicating that  many systems are probably not in a state of quasi-equilibrium, with their pulse periods still varying. An updated version of the diagram provided by \cite{Haberl2016}, which includes \sxp, is shown in Fig.\,\ref{fig:corbet}. The spin period is somewhat larger with respect to other sources with similar orbital periods, suggesting that the neutron star can still spin-up in the future. SXP\,1323 had, for example, a stable pulse period of around 1323\,s from 2000 to 2006, and then started to spin-up rapidly from 2006 to 2016 while keeping a constant flux \citep{Carpano2017}, to reach $\sim$1005\,s in 2019 November \citep{Haberl2019} and $\sim$977\,s in 2020 October \citep{2021arXiv210811517H}. 

The comparison of the system's long-term X-ray and OGLE-IV light curves indicates that the flux growth in both wavelength bands is correlated. The long-term optical variations of Be/X-ray binaries are believed to be related to the formation and depletion of the circumstellar disc around the donor star, giving rise to the Be phenomenon, as suggested in \cite{Rajoelimanana2011b}. A larger disc could lead to a higher accretion rate onto the neutron star and therefore increase its X-ray luminosity. Since \ero is the only instrument that has observed the source in its bright state with a sufficiently long exposure to retrieve the pulse period, new pointed observations would be necessary to explore the long-term impact of a larger Be disc on the pulse period derivative (a possible slow spin-up is suggested from these observations).

The optical spectrum is dominated by the \Halpha emission line emitted from the circumstellar disc surrounding the Be star and this line is therefore a good probe of the disc. The presence of the \Halpha emission line is one of the main classification criteria for Be/X-ray binaries. \cite{Coe2015, Antoniou2009} show that a strong correlation exists between the orbital period P$_\textrm{orb}$\,(d) and the line EW\,($\AA$). The linear best fit derived from the EW/P$_\textrm{orb}$ distribution of Be/X-ray pulsars in the SMC gives the relationship:

\begin{equation}
\textrm{H}_\alpha \textrm{EW}(\AA)=[-0.16 \times P_\textrm{orb}(\textrm{d})]-9.7 
\end{equation}
for orbital periods P$_\textrm{orb}\leq$ 150\,d \citep[][see their Fig.\,8]{Coe2015}.  
We note that the scatter observed by these latter authors in the EW/P$_\textrm{orb}$ distribution is large, especially for the short-period systems, and that it makes use of the largest EW ever measured.
The EW/P$_\textrm{orb}$ relationship is explained by the disc of the Be star being truncated by the neutron star during its orbit. The neutron star appears to act as a barrier, preventing the formation of an extended disc in systems with short orbital periods. This should also imply that the circumstellar disc, and in turn the \Halpha\,EW, are on average smaller in binary systems than for isolated Be stars. Correlations between the spectral parameters of the \Halpha line and rotational velocity have been observed in many Be stars and are interpreted as evidence for rotationally dominated circumstellar discs \citep{Dachs1986,Hanuschik1988, Reig2011}. A monitoring of the \Halpha\,EW and profile over a period of a few months could both confirm the 22.3\,d orbital period and provide constraints on the inclination angle of the Be disc.

\begin{acknowledgements}  
This work is based on data from \ero, the primary instrument aboard \srg, a joint Russian-German science mission supported by the Russian Space Agency (Roskosmos), in the interests of the Russian Academy of Sciences represented by its Space Research Institute (IKI), and the Deutsches Zentrum f{\"u}r Luft- und Raumfahrt (DLR). The \srg spacecraft was built by Lavochkin Association (NPOL) and its subcontractors, and is operated by NPOL with support from the Max Planck Institute for Extraterrestrial Physics (MPE).
The development and construction of the eROSITA X-ray instrument was led by MPE, with contributions from the Dr. Karl Remeis Observatory Bamberg \& ECAP (FAU Erlangen-N{\"u}rnberg), the University of Hamburg Observatory, the Leibniz Institute for Astrophysics Potsdam (AIP), and the Institute for Astronomy and Astrophysics of the University of T{\"u}bingen, with the support of DLR and the Max Planck Society. The Argelander Institute for Astronomy of the University of Bonn and the Ludwig Maximilians Universit{\"a}t Munich also participated in the science preparation for \ero.
The \ero data shown here were processed using the \texttt{eSASS}/\texttt{NRTA} software system developed by the German \ero consortium.
The optical spectrum presented in this paper were obtained with the Southern African Large Telescope (SALT).
\end{acknowledgements}

\bibliographystyle{aa}
\bibliography{article} 

\end{document}